# SOME NOTIONS ABOUT SEHR SPECTRA OF CRYSTAL VIOLET

A.M. Polubotko[1] V.P. Smirnov[2]

[1]A.F. Ioffe Physico-Technical Institute Russian Academy of Sciences, Politechnicheskaya 26 194021 Saint Petersburg RUSSIA E-mail: alex.marina@mail.ioffe.ru

[2]State University of Information Technologies, Mechanics and Optics, Kronverkskii 49, 197101 Saint Petersburg RUSSIA

## Abstract

The paper presents our opinion on the investigations and results of various kind of vibrational spectra of crystal violet (CV). After our analysis of the published spectra we came to conclusion, that the assignment of vibrations in this molecule remains questionable. Some experimental and calculations data permit us to conclude, that the vibrational bands in the SEHR spectra of CV contain the bands, caused by vibrations with $A_1$ and $A_2$ irreducible representations, that strongly confirm our idea about the main reason of surface enhanced optical processes and SEHRS in particular-the strong dipole and quadrupole light-molecule interactions.

Investigations of symmetry properties of the SEHR spectra of symmetrical molecules demonstrate existence of the bands, caused by unitary irreducible representations of corresponding symmetry groups, which are forbidden in usual hyper Raman spectra. This result can be well explained by our dipole-quadrupole theory of surface enhanced optical processes [1] and by the SEHRS theory [2-4]. In particular we analyzed the SEHR spectra of phenazine, pyrazine, trans-1, 2-bis (4-pyridyle) ethylene and pyridine. As it was demonstrated in [3,4] the SEHR spectra of the first two above mentioned molecules contain these forbidden lines, and can be completely explained by the dipole-quadrupole theory, while the details of investigation of other two spectra can be explained too. These details demonstrate the validity of our theory. One of the difficulties which prevent to demonstrate the validity of our theory for some symmetrical molecules is the absence of reliable assignment for large molecules. This situation arises because of a small distance between the wavenumbers of vibrations with various symmetry in large molecules, when this distance is significantly less or comparable with calculation errors of the used calculation methods. It refers to investigation of the SEHR spectrum of trans-1,2-bis (4-pyridyle) ethylene [3] first of all. Corresponding criticism is contained in [5]. Analogous situation arises for the analysis of the SEHR spectrum of CV. The most reliable assignment of vibrations was made in [6] where the molecule is referred to the $D_3$ symmetry group. In [7] IR, SER and SEHR spectra were partially investigated, however the authors considered that the molecule refers to the $C_3$ symmetry group. Recently the authors of [8,9] investigated SER, SEHR, usual Raman, IR, and hyper Raman spectra, considering that the molecule possesses by $D_3$ symmetry, such as in [6]. In addition they made their own assignment on the base of the DFT and B3LYP-311G methods. One should note, that careful consideration of experimental and calculated vibrational bands, vibrational wavenumbers, and possible assignment published in all these papers, reveal large discrepancy in calculated values and assignment of the bands (see [6-9]). Therefore our attempts to make our own assignment of the SEHR bands failed in fact and we consider, that the matter of precise assignment is spurious now. However there are several bands, which assigned to the $A_1$ irreducible representation in all these works. They are the bands at 1621 and 1389 $cm^{-1}$. In addition comparison of the IR, SER and SEHR spectra, published in [7] points out that the bands 1298, 1370, 1478 and 1588 $cm^{-1}$ are observed in all these spectra. Therefore in accordance with selection rules in IR spectra these bands can be caused only by vibrations with $A_2$ and $E$ symmetry. Consideration of the SER and SEHR spectra in accordance with the dipole-quadrupole theory [1] demonstrates, that these bands may be caused preferably by the $(Q_{main} - d_z)$ and $(Q_{main} - Q_{main} - d_z)$ contributions in SERS and

SEHRS respectively. This refer first of all to the bands at 1370 and 1589 $cm^{-1}$ because of their large relative intensity in SERS and SEHRS. The assignment of these bands to the $E$ symmetry may also be possible since the bands may be caused by the $(Q_{main} - (d_x, d_y))$ and $(Q_{main} - Q_{main} - (d_x, d_y))$ contributions, for molecules, superposed in the first and second layers having orientation which is not parallel to the surface. As it was considered earlier this fact depends on the coverage of substrate which is not known in [7]. Thus apparently there is a multilayer coverage in these experiments. Appearance and existence of the bands with $A_1$ symmetry in the SEHR spectra [6,8,9] is in agreement with our theory. One should note, that in the $D_3$ symmetry group, appearance of the bands, caused by vibrations with all possible irreducible representations in SEHRS formally is possible in a pure dipole theory, such as in pyridine. However appearance of the bands with $A_2$ symmetry in SERS can be explained only by the $(Q_{main} - d_z)$ type of the scattering. Therefore the correct explanation of the SEHR spectra of CV should be made on the base of the dipole-quadrupole theory, which is able to explain all other features of the SER and SEHR spectra of CV also. One should note, that the SEHR spectra of CV in the above mentioned works sometimes were obtained with the incident radiation, when the resonance scattering is possible. The resonance character of these processes must not change selection rules for contributions [3,4]. Therefore the result, that the bands, caused by vibrations with $A_1$ and $A_2$ symmetry are observed should be valid in the case of surface enhanced resonance Raman scattering and strongly support our theory.